\documentclass[12pt]{iopart}
\expandafter\let\csname equation*\endcsname\relax
\expandafter\let\csname endequation*\endcsname\relax
\usepackage{graphicx}
\usepackage{amsmath}
\usepackage{amssymb}
\usepackage{amsfonts}
\usepackage{booktabs}
\usepackage{overpic}
\usepackage{bm}
\usepackage{color}
\newcommand{\revision}[1]{{\color{black}#1}}
\usepackage{romannum}

\usepackage{fancyhdr} 
\usepackage[style=phys, sorting=none, backend=biber]{biblatex}
\begin{document}

\title{Linear-Stability-Guided Two-Stage Surrogate Model for Nonlinear ITG-TEM Turbulence}

\author{Y.W. Cho$^1$, C.G. Wan$^{2,*}$, Z.S. Qu$^2$, C.J. Ajay$^2$, K.P. Li$^2$, K. Lim$^2$, S.S. Kim$^1$, Y.-H. Lee$^1$, J.M. Kwon$^1$, and X. Garbet$^{2,3}$}

\address{$^1$Korea Institute of Fusion Energy, Daejeon, Republic of Korea}
\address{$^2$School of Physical and Mathematical Sciences, Nanyang Technological University, Singapore 637371, Singapore}
\address{$^3$CEA, IRFM, F-13108 Saint-Paul-lez-Durance, France}

\ead{chenguang.wan@ntu.edu.sg}

\begin{abstract}
We develop a two-stage surrogate model framework to predict the linear and nonlinear properties of ITG-TEM turbulence from GENE gyrokinetic simulations.
The first stage uses a hierarchical ensemble of stability and mode classifiers, along with global and mode-specific regressors, to predict the linear growth rate and real frequency.
The second stage uses these predictions as linear spectra in a Fourier neural operator (FNO)-based nonlinear surrogate model.
The nonlinear model predicts saturated potential and heat flux spectra, together with zonal flow quantities.
Two linear drive scalars, the maximum and integral of the linear growth rate over a low-wavenumber range, are also supplied directly to the networks predicting scalar quantities.
For the architecture studied here, comparisons with models omitting either or both input paths show that linear information can improve spectral and scalar predictions under extrapolation toward stronger linear drive.
\end{abstract}

\noindent{\it Keywords}: Surrogate Model, Gyrokinetic Simulation, ITG-TEM Instability

\section{Introduction}
\label{sec:intro}

Reliable prediction of plasma transport is important for integrated modeling, scenario development, and real-time control of magnetically confined fusion plasmas. 
Integrated modeling frameworks combine magnetic equilibrium, kinetic profiles, heating and current drive, and transport within a common workflow.
Among these components, the repeated evaluation of turbulent transport can require a substantial fraction of the total computational cost. 
A sufficiently fast model of turbulent transport is therefore needed for applications that require many evaluations during optimization or time-dependent simulation.

Turbulent transport in the core plasma is commonly studied using gyrokinetic simulations. 
Local flux-tube codes such as GENE~\cite{jenko2000gene}, GKW~\cite{peeters2009gkw}, CGYRO~\cite{candy2016cgyro}, and GX~\cite{mandell2024gx} can resolve the kinetic instabilities and nonlinear interactions that determine the transport, but nonlinear simulations remain too expensive for direct use in iterative or real-time calculations. 
For instance, one nonlinear ion temperature gradient (ITG) and trapped electron mode (TEM) turbulence simulation with GENE requires approximately 8 A100 GPU-hours, compared with approximately 0.25 GPU-hours for the adiabatic-electron GX simulations used in our previous ITG database~\cite{wan2025multifidelity}. 
This cost makes the generation of a large nonlinear ITG-TEM database difficult and motivates a more data-efficient modeling strategy.

Machine learning surrogate models provide one approach for reducing the cost of turbulent transport prediction. 
Early studies trained neural networks to reproduce the fluxes of the quasilinear transport models QuaLiKiz and TGLF, enabling their use in real-time and integrated modeling calculations~\cite{citrin2015realtime,meneghini2017corepedestal,van2020fast}. 
For example, TORAX includes a QuaLiKiz-trained neural network surrogate model as an option for computing turbulent transport in time-dependent plasma profile simulations~\cite{citrin2024torax}.
To increase the fidelity of this approach, Citrin et al. constructed a quasilinear GENE database using linear gyrokinetic calculations and a saturation rule calibrated against dedicated nonlinear GENE simulations~\cite{citrin2023higherfidelity}. 
Related developments include a surrogate model for the linear eigenvalues obtained from QuaLiKiz and active-learning methods that reduce the number of transport calculations required to construct the training database~\cite{fransson2023qualikiz,zanisi2024adept}. 
More recently, TGLF-WINN has used the wavenumber-resolved flux contributions of TGLF as additional training information and combined this approach with Bayesian active learning~\cite{cao2026tglfwinn}. 
These models accelerate reduced quasilinear transport models, rather than learning directly from nonlinear gyrokinetic simulations.

Surrogate models based on nonlinear gyrokinetic data remain less common. 
\revision{A multi-fidelity surrogate model combines a large database of GPU-based GX simulations with a smaller database of higher-fidelity GKW simulations for ITG turbulence with adiabatic electrons~\cite{wan2025multifidelity}.}
A recent spatiotemporal model has also learned the evolution of potential fields and transport fluxes directly from nonlinear CGYRO simulations~\cite{ashourvan2026gkfieldflow}. 

More broadly, theory-guided data science provides a framework for incorporating scientific knowledge through the model architecture, learning procedure, or combination of theory-based and data-driven components~\cite{karpatne2017theory}. 
Physics-informed machine learning has also been applied to reduced turbulence and closure problems by incorporating physical constraints or model structure into the learning process~\cite{karniadakis2021physics,li2025pinn}. 
Physics-guided learning methods have also incorporated predictions from simplified physical models as additional features of data-driven models~\cite{daw2017pgnn,pawar2021physics}. 
Linear gyrokinetic information has also been used in multi-fidelity Gaussian process regression to improve predictions of experimentally inferred turbulent diffusion coefficients in JET. 
A related application combined quasilinear predictions with nonlinear gyrokinetic simulation results to improve particle flux predictions for an LHD discharge~\cite{maeyama2024multifidelity}.
A closely related approach has recently been applied to gyrokinetic turbulent transport, where $k_y$-resolved linear quantities were mapped to nonlinear saturated potential amplitudes within a quasilinear saturation model~\cite{sar2026sat3nn}. 
These studies demonstrate the potential of combining physical models with data-driven surrogate models for plasma turbulence. 
For nonlinear ITG-TEM surrogate modeling, the cost of simulations with kinetic electrons limits the available training data.
Predicting the spectral response also requires reproducing the distribution of potential fluctuations and heat flux across wavenumber, rather than only their integrated magnitudes.

In this work, we address this problem using a two-stage framework that combines linear and nonlinear GENE simulations. 
We first generate a large linear ITG-TEM database and train a linear surrogate model to predict stability, the dominant instability type, the linear growth rate, and the real frequency. 
The linear surrogate model is then used to restrict nonlinear database generation to clearly unstable conditions and to provide physics-informed inputs to the nonlinear surrogate model. 
The growth rate and real frequency are used as physical descriptors of the instability, rather than as low-fidelity estimates of the nonlinear targets.
The nonlinear surrogate model learns from nonlinear GENE data without imposing a quasilinear flux closure or adding governing equations to the loss function.

The nonlinear surrogate model uses a Fourier Neural Operator (FNO)~\cite{li2021fno} to predict saturated turbulence spectra and zonal flow quantities.
Linear stability information is supplied in two forms: the growth rate and real frequency at each $k_y$, referred to as the linear spectra, and the maximum and integrated growth rates over $k_y\leq1$, referred to as the linear drive scalars.

We compare four variants with different combinations of these inputs.
The variants show comparable performance within the training domain, but their differences become clearer under extrapolation toward stronger linear drive.
In this architecture, the linear drive scalars improve the overall spectral magnitude and zonal flow predictions, while adding the linear spectra improves the short-wavelength spectral predictions.

The remainder of this paper is organized as follows. 
Sec.~\ref{sec:linear} describes the generation of the linear ITG-TEM database and the development and validation of the linear surrogate model. 
Sec.~\ref{sec:nonlinear} presents the nonlinear database, the architecture of the nonlinear surrogate model, and the evaluation of the two linear information paths. 
Finally, Sec.~\ref{sec:conclusion} summarizes the main results and discusses future extensions of the framework.

\section{Development of linear surrogate model}
\label{sec:linear}

Nonlinear ITG--TEM simulations with kinetic electrons are too expensive to sample the full input parameter space directly. 
We therefore first construct a large linear database, for which each simulation is substantially less expensive, and use it to train a linear surrogate model. 
The model determines whether a parameter point is linearly unstable, identifies the dominant ITG or TEM branch, and predicts the corresponding linear growth rate and real frequency. 
This provides an efficient way to select unstable cases for the nonlinear database and supplies linear stability information to the nonlinear surrogate model described in Sec.~\ref{sec:nonlinear}.

\subsection{Generation of linear database}

The linear database is generated using the GENE gyrokinetic code~\cite{jenko2000gene}. 
All simulations are performed in a local flux tube with concentric circular magnetic flux surfaces. 
Both the linear and nonlinear GENE simulations are performed under collisionless conditions with equal ion and electron temperatures, $T_i=T_e$.
Each linear simulation uses a numerical resolution of $(N_{k_x},N_{k_y},N_z,N_{v_\parallel},N_\mu)=(16,1,16,48,16)$.
These quantities denote the numbers of radial and binormal Fourier modes, parallel grid points, parallel-velocity grid points, and magnetic-moment grid points, respectively.
Each simulation requires approximately 10 minutes on one NVIDIA A100 GPU.

The database spans the seven input variables listed in Table~\ref{tab:linear-scan}. 
Most cases are generated on a regular grid, while supplementary Latin hypercube sampling (LHS) cases~\cite{mckay1979lhs} improve the coverage between the regular grid points. 
The lower $k_y$ range is sampled more finely because the dominant ion scale growth rate structure and the transitions between stable, ITG, and TEM regimes are concentrated in this range. 
The higher $k_y$ range is sampled more coarsely to limit the total computational cost.

\begin{table}[t]
\centering
\caption{Input parameters of the linear GENE database. The continuous ranges are used for the supplementary LHS cases, and the listed values are used for regular sampling.}
\begin{tabular}{l c l}
\hline
Parameter & LHS range & Regular sampling values \\
\hline
$k_y\rho_s$          & $0.1\sim4.0$    & $0.1\sim1.0$ (step 0.1), $1.5\sim3.0$ (step 0.5) \\
$\varepsilon=r/R$    & $0.1\sim0.3$    & 0.07, 0.13, 0.17, 0.20, 0.26 \\
$q$                  & $1.0\sim3.0$    & 0.66, 1.0, 1.5, 2.0 \\
$\hat{s}$            & $-0.5\sim1.0$ & $-0.5$, 0.1, 0.4, 0.7 \\
$R/L_{T_i}$, $R/L_{T_e}$ & $2.0\sim8.0$ & 2, 4, 5, 6, 7, 8 \\
$R/L_n$              & $0.5\sim3.0$    & 1.0, 1.5, 2.0, 2.5, 3.0 \\
\hline
\end{tabular}
\label{tab:linear-scan}
\end{table}

Approximately 115,000 cases are obtained from the regular scan, and 4,000 additional cases are obtained through LHS. 
\revision{Screening excludes extreme-frequency outputs and selected zero-output cases with $\gamma=\omega=0$ that disagree with both the global and mode-routed predictions of an earlier surrogate model.
Approximately 102,000 cases remain for training and evaluation.}
The total computational cost of generating the linear database is approximately $2\times10^4$ A100 GPU hours.

For each input vector $\mathbf{x}=(k_y\rho_s,\varepsilon,q,\hat{s},R/L_{T_i},R/L_{T_e},R/L_n)$, the target quantities are the linear growth rate $\gamma$ and the corresponding real frequency $\omega$ obtained from the GENE simulation.
Here, $q$ is the safety factor, $\hat{s}$ is the magnetic shear, $R/L_{T_i}$ is the normalized ion temperature gradient, $R/L_{T_e}$ is the normalized electron temperature gradient, and $R/L_n$ is the normalized density gradient.
$\gamma$ and $\omega$ are normalized by $c_s/R$, where $c_s=\sqrt{T_{\mathrm{ref}}/m_i}$ is the reference sound speed, $T_{\mathrm{ref}}$ is the reference temperature, $m_i$ is the deuterium ion mass, and $R$ is the major radius.
$\rho_s=c_s/\omega_{ci}$ is the reference Larmor radius, where $\omega_{ci}=|e|B/m_i$ is the ion cyclotron frequency, $e$ is the electron charge, and $B$ is the reference magnetic field.
The binormal wavenumber is normalized by $\rho_s^{-1}$. 
For simplicity, the dimensionless wavenumber $k_y\rho_s$ is denoted by $k_y$ in the rest of this paper.
A case is labeled as stable when $\gamma\leq0$. 
An unstable case is labeled as ITG when $\omega>0$ and as TEM when $\omega<0$, following the frequency sign convention used in the GENE postprocessing.

\subsection{Framework of linear surrogate model}

The linear surrogate model maps the seven input parameters to the stability, dominant instability type, growth rate, and real frequency. 
Synthetic stable points at $k_y\rho_s=0$, with $\gamma=\omega=0$, are included in the training set to impose the expected boundary behavior at zero wavenumber.

\begin{figure}[t]
\centering
\begin{overpic}[width=0.90\textwidth]{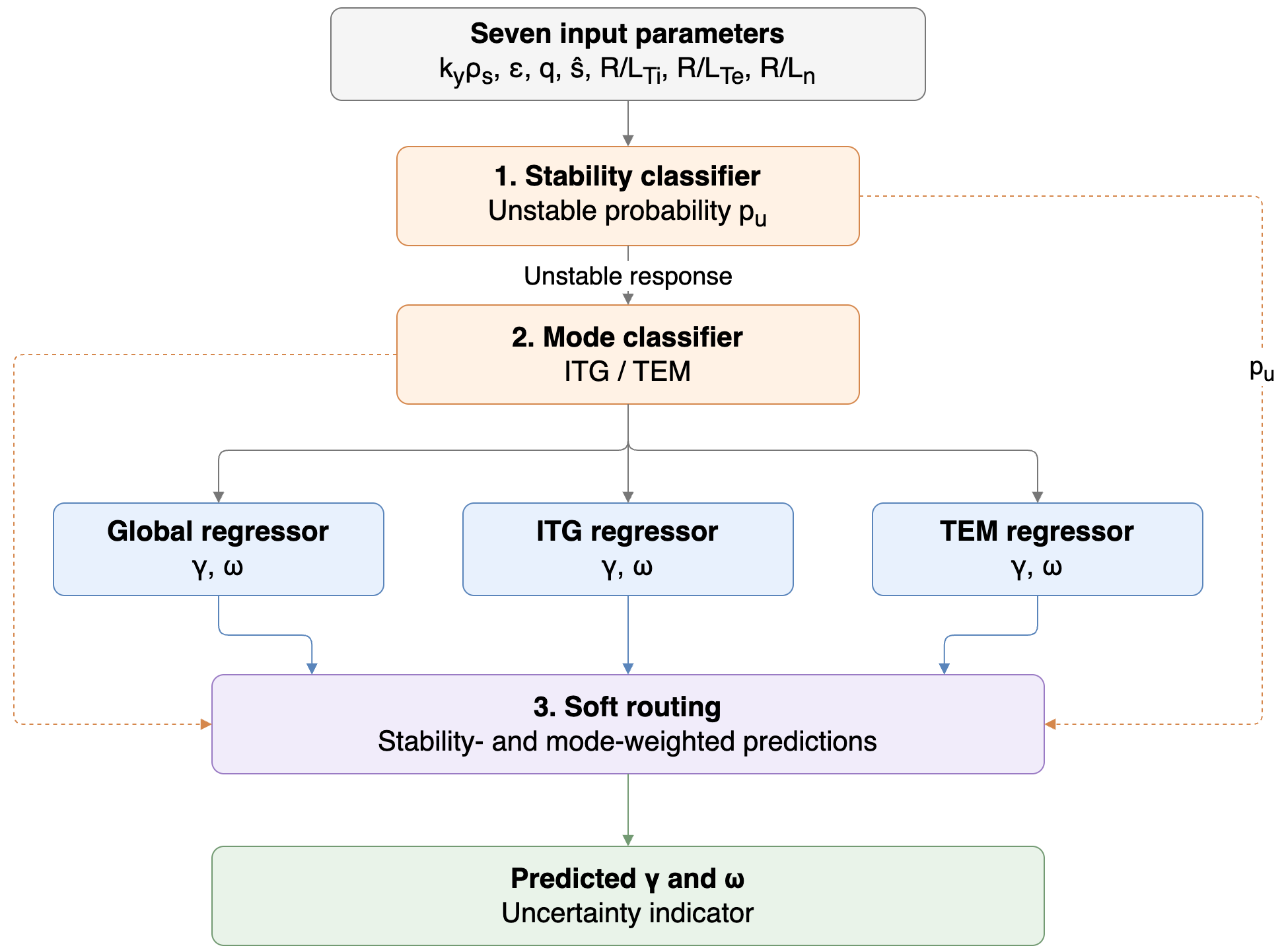}
\end{overpic}
\caption{Workflow of the linear surrogate model. Solid arrows indicate the prediction stages, and dashed arrows carry classifier probabilities for soft routing.}
\label{fig:lin-framework}
\end{figure}

The model uses a hierarchical ensemble architecture, as illustrated in Fig.~\ref{fig:lin-framework}. 
A binary classifier first predicts the unstable probability $p_{\mathrm{u}}(\mathbf{x})$. 
For the unstable response, a second classifier predicts the probabilities $p_m(\mathbf{x})$ of the ITG and TEM
branches. 
Mode-specific regressors predict the response associated with each branch, while a global regressor predicts the same quantities using all unstable cases without separating the instability type.

\revision{Each classifier and regressor uses an ensemble of seven multilayer perceptrons (MLPs), each trained on a different random sample of the training data.
Sampling allows each case to be selected more than once~\cite{efron1979bootstrap}.}
The MLPs have hidden dimensions $(256,256,128,64)$ and use layer normalization, SiLU activation functions, and a dropout rate of 0.1. 
The classifiers are trained using cross entropy, and the regressors are trained using a mean squared error objective. 
For the present framework, the relevant regression outputs are $\gamma$ and $\omega$.

The component predictions are weighted by the classification probabilities, a procedure called soft probabilistic routing.
For $A\in\{\gamma,\omega\}$, the final prediction is
\begin{equation}
\widehat{A}(\mathbf{x})
=p_{\mathrm{u}}(\mathbf{x})
\left[
\alpha(\mathbf{x})\widehat{A}_{\mathrm{g}}(\mathbf{x})
+\left(1-\alpha(\mathbf{x})\right)
\sum_{m\in\{\mathrm{ITG},\mathrm{TEM}\}}
p_m(\mathbf{x})\widehat{A}_m(\mathbf{x})
\right],
\label{eq:linear-soft-routing}
\end{equation}
where $\widehat{A}_{\mathrm{g}}$ and $\widehat{A}_m$ denote the global and mode-specific predictions, respectively. The mixing coefficient is the normalized entropy of the mode probabilities,
\begin{equation}
\alpha(\mathbf{x})
=-\frac{\displaystyle\sum_m p_m(\mathbf{x})\log p_m(\mathbf{x})}
{\log 2}.
\label{eq:linear-routing-entropy}
\end{equation}
The global regressor therefore receives more weight when the ITG and TEM classification is uncertain, whereas the mode-specific regressors receive more weight when one branch is identified clearly. 
Multiplication by $p_{\mathrm{u}}$ also connects the unstable prediction continuously to the stable value. 
This procedure avoids discontinuous changes at the stability and ITG--TEM classification boundaries. 
The dispersion among the ensemble members within each component is propagated through the routing procedure and is used as an uncertainty indicator~\cite{lakshminarayanan2017deepensemble}.

\subsection{Training and validation of linear surrogate model}

The screened database is divided into training, validation, and test sets with a ratio of 70:15:15 while preserving the stable and unstable proportions. 
The supplementary LHS cases are included before this split rather than being used as a separate validation set. 
The held-out test set contains 15,365 cases and is used only for the results reported in this subsection.

\begin{figure}[t]
\centering
\includegraphics[width=0.90\textwidth]{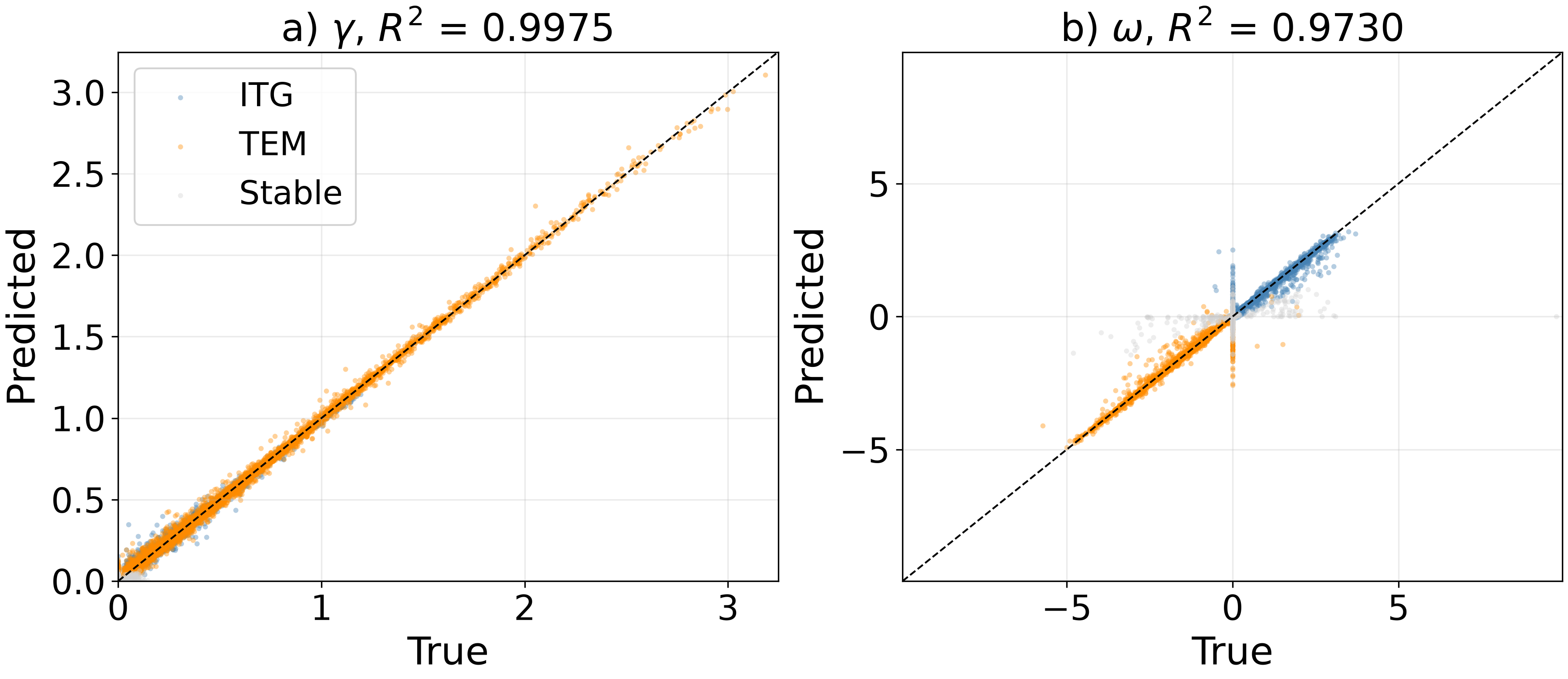}
\caption{Parity plots for a) the linear growth rate $\gamma$ and b) the real frequency $\omega$. Colors indicate the predicted mode, with blue for ITG, orange for TEM, and gray for stable cases.}
\label{fig:lin-r2}
\end{figure}

Fig.~\ref{fig:lin-r2} compares the predicted growth rates and frequencies with the corresponding GENE results. 
The model gives $R^2=0.9975$ for $\gamma$ and $R^2=0.9730$ for $\omega$. 
The remaining errors are concentrated mainly near the stability and ITG--TEM transition boundaries, where small changes in the inputs can change the assigned branch.

\begin{figure}[t]
\centering
\includegraphics[width=0.98\textwidth]{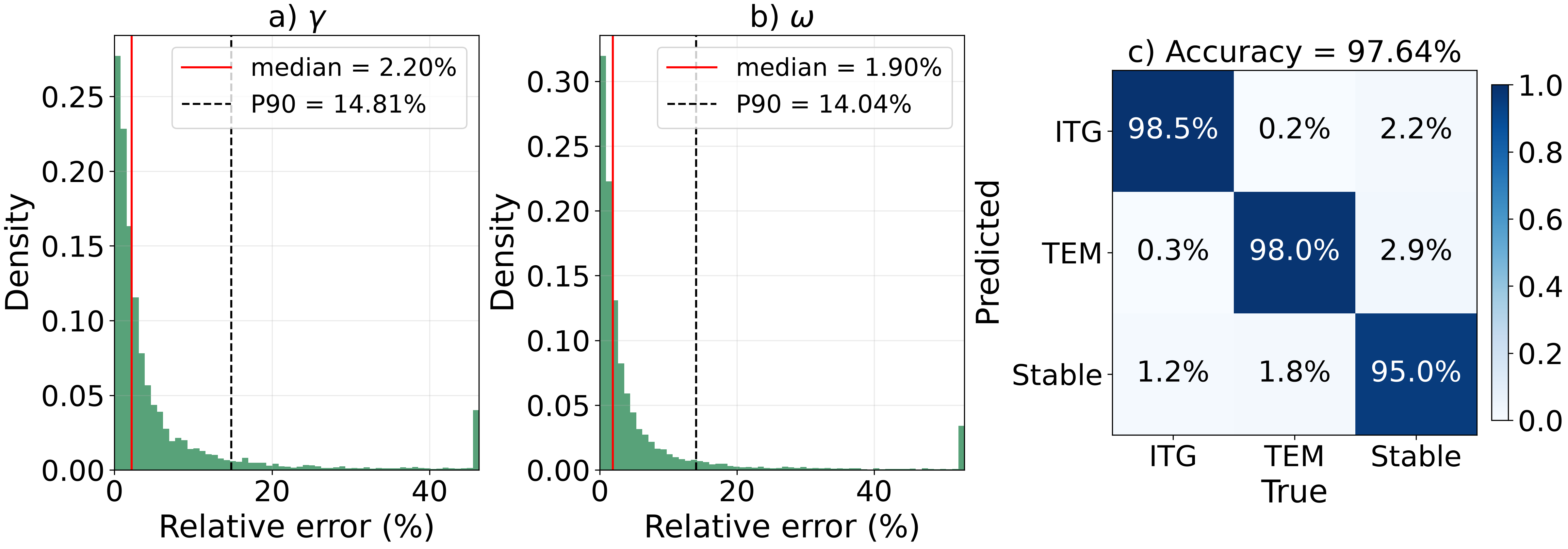}
\caption{Relative error distributions of a) the growth rate and b) the real frequency predictions, and c) the stability and mode classification accuracy of the linear surrogate model.}
\label{fig:lin-error-acc}
\end{figure}

For unstable test cases with nonzero target magnitude, the median absolute relative errors of $\gamma$ and $\omega$ are 2.20\% and 1.90\%, respectively, and their 90th percentile errors remain below 15\%. 
The overall stability and mode classification accuracy is 97.64\%\revision{, as shown in Fig.~\ref{fig:lin-error-acc}(c)}.
Most classification errors occur close to the stability or mode transition boundaries. 
The corresponding growth rates are generally small, which limits their effect on the subsequent selection of strongly unstable nonlinear cases.

\begin{figure}[t]
\centering
\begin{overpic}[width=0.98\textwidth]{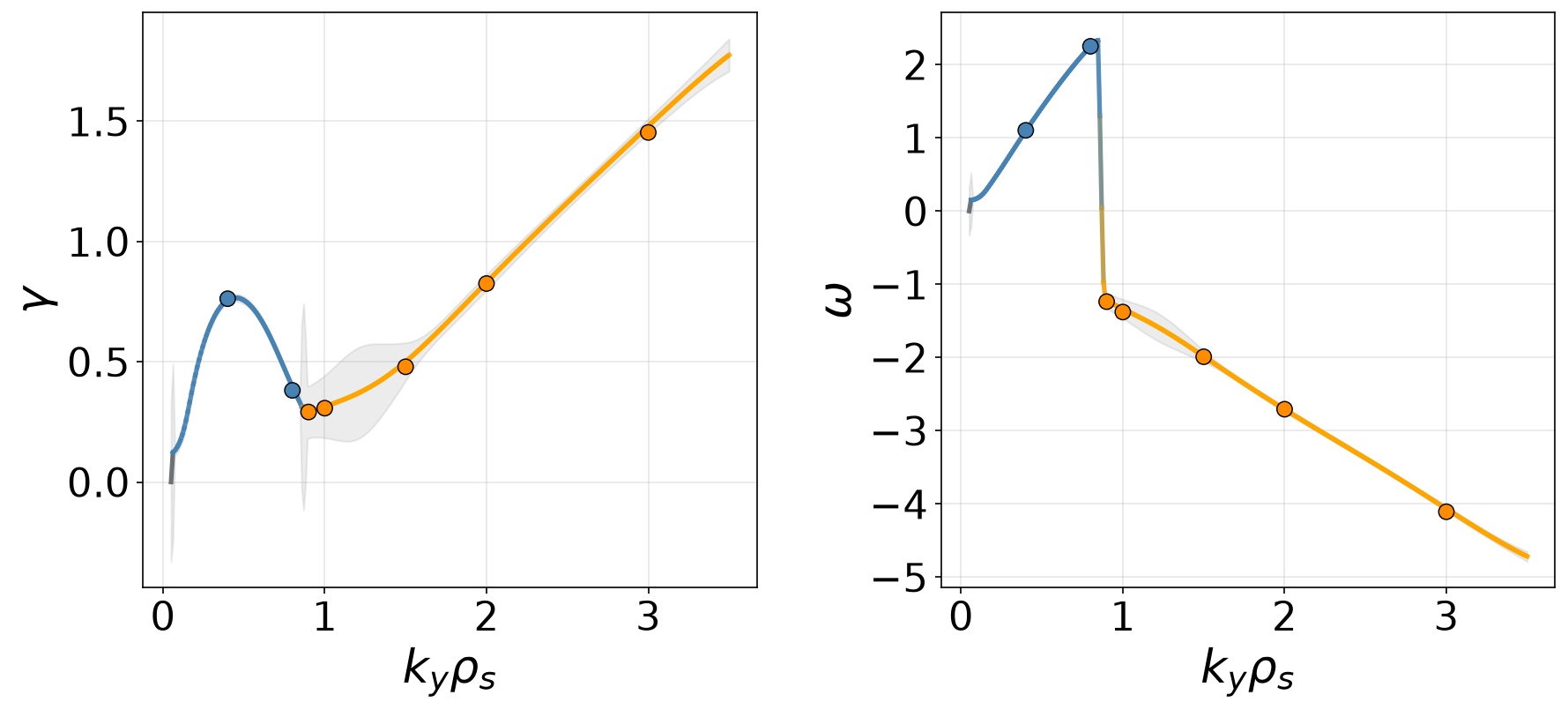}
\put(12,40){\textbf{a)}}
\put(63,12){\textbf{b)}}
\end{overpic}\\
\begin{overpic}[width=0.98\textwidth]{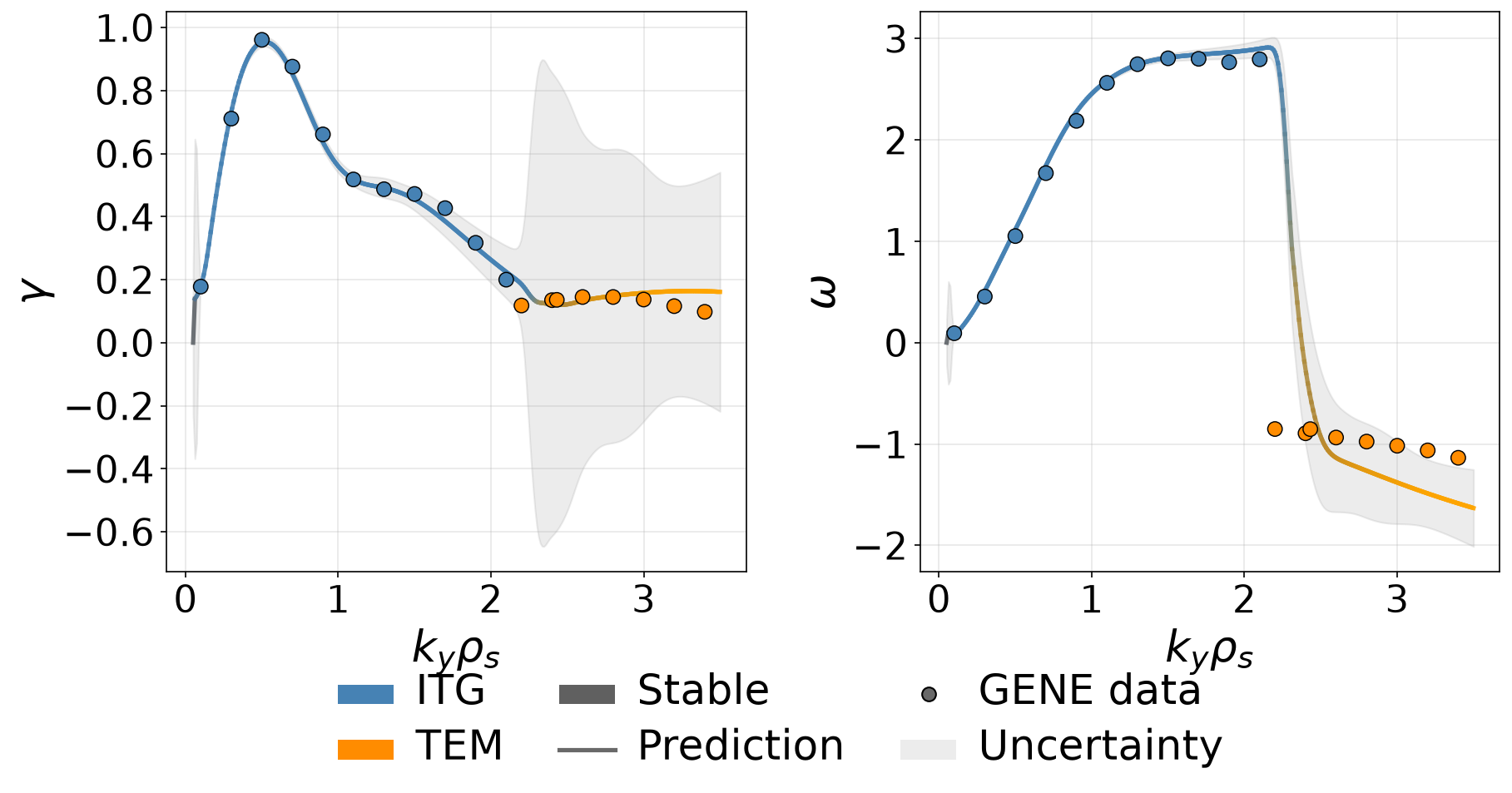}
\put(14,22){\textbf{c)}}
\put(64,22){\textbf{d)}}
\end{overpic}
\caption{Example linear growth rate (left) and real frequency (right) predictions. For $(\varepsilon,q,\hat{s},R/L_{T_i},R/L_{T_e},R/L_n)$, panels a) and b) use the training case $(0.17,1.5,0.7,8.0,7.0,1.5)$, and panels c) and d) use the \revision{case not used for training} $(0.22,1.92,-0.27,7.93,3.53,2.31)$. Dark gray denotes stable predictions. Light gray bands indicate combined ensemble dispersion, not calibrated confidence intervals or output-specific standard deviations.}
\label{fig:lin-examples}
\end{figure}

Representative $k_y$ spectra are shown in Fig.~\ref{fig:lin-examples}. 
The first case is located on the regular sampling grid and is included in the training data.
\revision{The second case was not used for training.}
The model reproduces the growth rate and frequency over most of the resolved range in both cases.
The uncertainty increases near the mode transition and in the more sparsely sampled higher $k_y$ range, but the main spectral structure and the transition between ITG and TEM branches are retained. 
These results show that the linear surrogate model provides sufficiently accurate and continuous linear stability information for the nonlinear database selection and nonlinear surrogate model developed in Sec.~\ref{sec:nonlinear}.

\section{Development of nonlinear surrogate model}
\label{sec:nonlinear}

\subsection{Generation of nonlinear database}

The nonlinear surrogate model is developed to predict saturated potential fluctuation and heat flux spectra for parameter scans in which direct nonlinear GENE simulations are too expensive~\cite{jenko2000gene}. 
Nonlinear ITG-TEM simulations are considerably more costly than the linear calculations used in Sec.~\ref{sec:linear}. 
Each nonlinear case is computed in concentric circular geometry with a numerical resolution of $(N_{k_x},N_{k_y},N_z,N_{v_\parallel},N_\mu)=(128,48,16,48,16)$ and is evolved to $t=50\sim100\,R/c_s$.
A typical case uses two A100 GPUs for approximately 4 hours, corresponding to about 8 A100 GPU-hours.

The averaging window is selected from the ion heat flux time series.
After the initial nonlinear saturation, an additional interval equal to 15\% of the total recorded duration is excluded to avoid transient effects. 
The saturated quantities are then averaged from the time when the smoothed heat flux becomes approximately stationary to the end of the recorded data.

The six input parameters are sampled over the ranges in Table~\ref{tab:nonlinear-scan} using Latin hypercube sampling~\cite{mckay1979lhs}. 
To avoid spending computational resources on weakly unstable or stable conditions, the linear surrogate model is used to screen the sampled inputs before the nonlinear simulations are performed. 
A case is retained for nonlinear simulation when its maximum predicted growth rate satisfies $\gamma_{\max}>0.5$ and more than approximately 40\% of the sampled $k_y$ modes are predicted to be unstable.
This procedure deliberately restricts the nonlinear database to clearly unstable ITG-TEM conditions.
The resulting database contains 660 nonlinear GENE cases with confirmed nonlinear saturation and is used for all model comparisons.

The nonlinear postprocessing provides two-dimensional potential amplitude and total heat flux spectra over $(k_x,k_y)$. 
Here, we define the total heat flux as the sum of the ion and electron contributions, $Q_{\mathrm{tot}}=Q_i+Q_e$.
These outputs are reduced to the spectral quantities described in Sec.~\ref{sec:nl-model} before they are used as training targets.
Unless stated otherwise, the heat flux is normalized by $Q_{\mathrm{ref},R}=n_{\mathrm{ref}}T_{\mathrm{ref}}c_s(\rho_s/R)^2$, following the native GENE normalization with reference length $R$.
Likewise, the electrostatic potential is normalized by $\phi_{\mathrm{ref},R}=(T_{\mathrm{ref}}/e)(\rho_s/R)$.
For the total heat flux scatter plots, the values are converted to minor-radius-based gyro-Bohm units assuming $a/R=1/3$.

\begin{table}[t]
\caption{Scan ranges of the input parameters and wavenumbers used for the nonlinear GENE database. The $k_y\rho_s$ interval is 0.04.}
\centering
\begin{tabular}{lcccccc}
\toprule
Parameter & $k_y\rho_s$ & $\varepsilon=r/R$ & $q$ & $\hat{s}$ & $R/L_{T_i}$, $R/L_{T_e}$ & $R/L_n$ \\
\midrule
Range & $0\sim1.88$ & $0.1\sim0.3$ & $1.0\sim3.0$ & $0.4\sim1.0$ & $2.0\sim8.0$ & $0.5\sim3.0$ \\
\bottomrule
\end{tabular}
\label{tab:nonlinear-scan}
\end{table}

\subsection{Framework of nonlinear surrogate model}
\label{sec:nl-model}

\begin{figure}[t]
\centering
\begin{overpic}[width=0.98\textwidth]{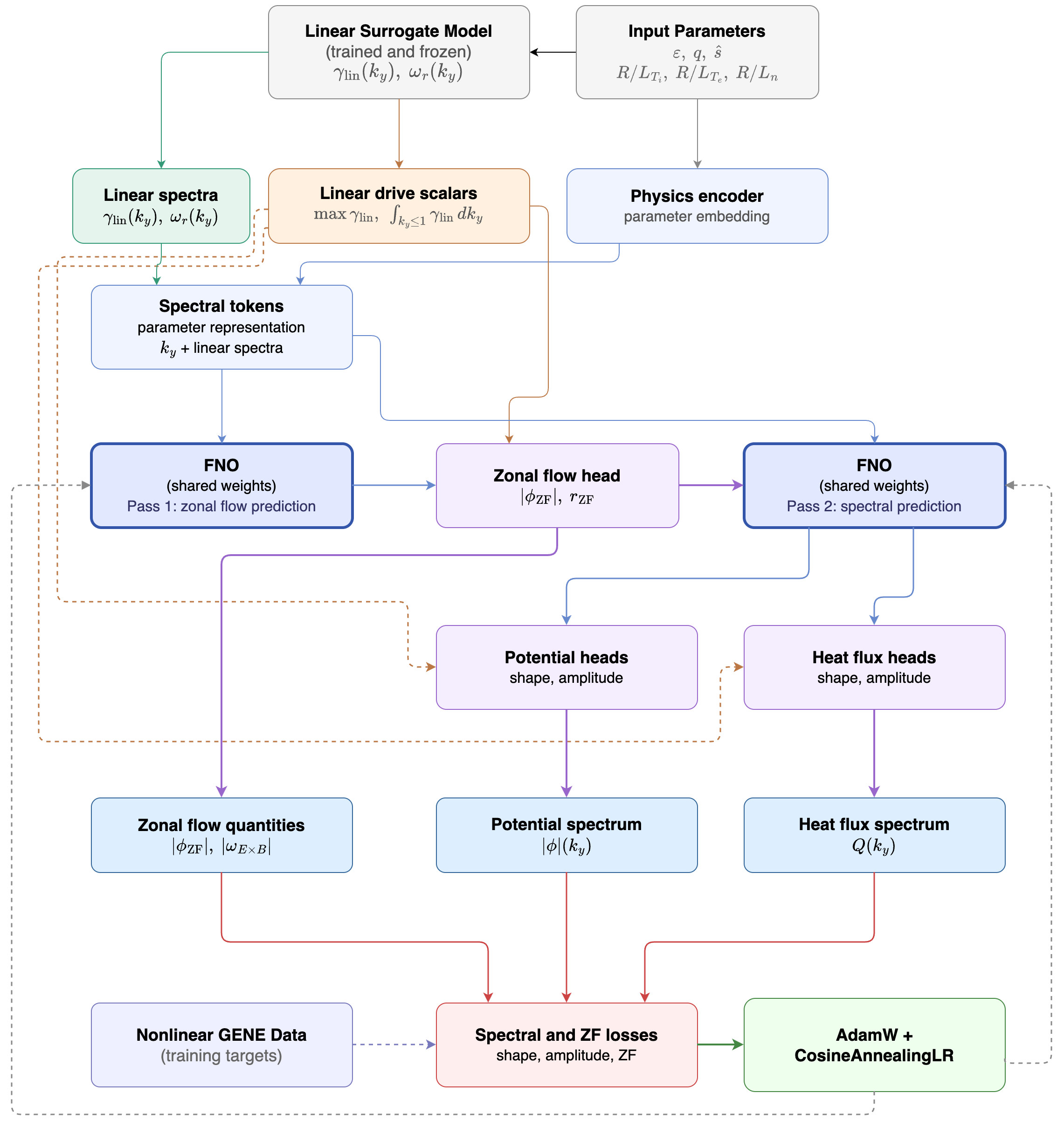}
\end{overpic}
\caption{Overview of the nonlinear surrogate model. Green and orange branches denote the independently switched linear spectra and linear drive scalar inputs. The two FNO passes share weights. \revision{The potential and heat flux blocks show the spectral shape and integrated amplitude heads.} Some direct inputs to the scalar heads are omitted for clarity.}
\label{fig:nl-strategy}
\end{figure}

Fig.~\ref{fig:nl-strategy} shows the overall structure of the nonlinear surrogate model.
The inputs are the six equilibrium and gradient parameters $(\varepsilon,q,\hat{s},R/L_{T_i},R/L_{T_e},R/L_n)$, together with the linear growth rate and real frequency predicted by the linear surrogate model.
From these inputs, the model predicts saturated potential and heat flux spectra over 47 nonzero $k_y$ modes, together with zonal flow quantities at $k_y=0$.

At the input stage of Fig.~\ref{fig:nl-strategy}, a multilayer perceptron (MLP) with two hidden layers, called the physics encoder, transforms the six normalized input parameters into a shared representation of the plasma conditions.
The linear surrogate model supplies the growth rate $\gamma_{\mathrm{lin}}(k_y)$ and real frequency $\omega_r(k_y)$ at each nonzero $k_y$ location.
These growth rate and frequency spectra are referred to as the linear spectra.
For each nonzero $k_y$ mode, the shared parameter representation is concatenated with its $k_y$, growth rate, and frequency.
A shared projection network maps these combined inputs into a spectral token.
In parallel, the maximum growth rate and the integral of $\gamma_{\mathrm{lin}}$, both evaluated over $k_y\leq1$, are calculated as the linear drive scalars.
The maximum measures the strongest growth in this range, while the integral also reflects the extent of the unstable response across $k_y$.
The low $k_y$ range is chosen to represent the drive relevant to the principal nonlinear spectral peak and zonal flow regulation.
Zonal flows play an important role in ITG saturation, whereas their role in TEM turbulence depends on the wavelength and drive regime~\cite{garbet2010gyrokinetic,ernst2009zonal,chen2022zonal}.
Nonlinear transport spectra can also peak below the wavenumber of the fastest growing linear mode~\cite{dannert2005tem}.

In the first pass, the FNO~\cite{li2021fno} processes the spectral tokens together with a zonal flow token to obtain a representation for zonal flow prediction.
Its spectral convolutions exchange information across the token sequence.
This representation is concatenated with the parameter representation and the linear drive scalars and supplied to a separate output network, called the zonal flow head.
The head predicts the root-mean-square (RMS) potential amplitude $|\phi_{\mathrm{ZF}}|$ and the ratio $r_{\mathrm{ZF}}=|\omega_{E\times B}|/|\phi_{\mathrm{ZF}}|$, where $|\omega_{E\times B}|\sim|k_x^2\phi_{\mathrm{ZF}}|$ is $E\times B$ shearing rate.
This separates the overall zonal amplitude from a radial structure factor associated with an effective $k_x^2$ scale.

For the second pass, the predicted zonal flow amplitude and the shearing rate reconstructed from this ratio are inserted into the zonal flow token.
The same FNO processes the updated token together with the original nonzero $k_y$ input features, allowing the spectral predictions to incorporate the predicted zonal flow.
The resulting spectra are not fed back for a further zonal flow prediction.

\revision{The lower output branches in Fig.~\ref{fig:nl-strategy} predict the potential and heat flux spectra.}
The potential and heat flux spectra are reduced over the sampled positive $k_x$ modes before their $k_y$ dependence is modeled. 
We denote the number of retained positive radial modes by $N_{k_x}^{+}=63$, distinct from the full simulation resolution $N_{k_x}=128$.
The potential spectrum is defined by the discrete RMS amplitude
\begin{equation}
|\phi|(k_y)
=
\left[
\frac{1}{N_{k_x}^{+}}
\sum_{j=1}^{N_{k_x}^{+}} |\phi(k_{x,j},k_y)|^2
\right]^{1/2},
\label{eq:nl-kx-rms}
\end{equation}
whereas the heat flux spectrum is defined by
\begin{equation}
Q(k_y)
=
\frac{1}{N_{k_x}^{+}}
\sum_{j=1}^{N_{k_x}^{+}} Q_{\mathrm{tot}}(k_{x,j},k_y).
\label{eq:nl-kxavg-Q}
\end{equation}
The total electrostatic energy and heat flux are defined as $N_{k_x}^{+}\sum_{k_y>0}[|\phi|(k_y)]^2$ and $N_{k_x}^{+}\sum_{k_y>0}Q(k_y)$, respectively.

The potential and heat flux spectra are predicted using separate heads for the spectral shape and integrated amplitude.
For each spectrum $A\in\{|\phi|,Q\}$, the shape head predicts the normalized spectral shape $S_A(k_y)$, and the amplitude head predicts its integrated amplitude $T_A$.
These quantities are defined as
\begin{align}
S_A(k_y) &=\frac{A(k_y)}{\int A(k_y')\,d k_y'},
\label{eq:nl-shape}\\
T_A &=\int A(k_y)\,d k_y.
\label{eq:nl-amplitude}
\end{align}
The complete predicted spectrum is reconstructed as
\begin{equation}
A_{\mathrm{pred}}(k_y) =T_{A,\mathrm{pred}}S_{A,\mathrm{pred}}(k_y).
\label{eq:nl-spectrum}
\end{equation}
The integrated amplitude refers to the $k_y$ integral of the reduced spectrum, whereas the mode-summed totals shown later are calculated from the reconstructed spectra.
In particular, the integrated potential amplitude is distinct from the total nonzonal electrostatic energy, which depends on the squared potential amplitude.

\revision{The shape heads use the nonzero $k_y$ feature vectors from the second FNO pass.}
The integrated amplitude heads use the physics encoder output, predicted zonal flow quantities, and linear drive scalars.
The FNO has two layers, with an embedding width of 16, a feed-forward dimension of 32, and 16 Fourier modes.
Each output head has one hidden layer with a SiLU activation.
\revision{The potential and heat flux shape heads and the zonal flow head use a hidden width of 32, while the two integrated amplitude heads use a hidden width of 16.}

\revision{The spectral shapes, integrated amplitudes, and zonal flow quantities are supervised separately.}
A small smoothness penalty is also included to suppress isolated spikes between neighboring $k_y$ modes.
Gradients from the integrated amplitude heads are detached along the zonal flow input path, so their losses do not update the zonal flow prediction through this connection.
The model is trained using the AdamW optimizer with a cosine annealing learning rate schedule.
Training proceeds with a maximum of 300 epochs and early stopping after 60 epochs without improvement in the validation loss.
The output parameterizations and loss formulation are given in \ref{sec:nl-training-details}.

\subsection{Evaluation of linear stability information in nonlinear surrogate model}
\label{sec:nl-eval}

In this subsection, we evaluate how the two forms of linear stability information introduced in Sec.~\ref{sec:nl-model} contribute to the nonlinear surrogate model. 
The first consists of the linear spectra of growth rate and real frequency supplied to the spectral tokens. 
The second consists of the linear drive scalars supplied directly to the zonal flow and integrated amplitude heads. 
Four model variants are constructed by retaining or removing these two input paths, as summarized in Table~\ref{tab:nl-variants}. 
All four variants receive the same six input parameters. 
The variant names indicate which additional linear information is supplied.
All variants otherwise use the same network architecture, loss function, and training procedure.

\begin{table}[t]
\centering
\caption{Linear stability information supplied to each model variant.}
\begin{tabular}{lcc}
\toprule
Variant
& Linear spectra
& Linear drive scalars \\
\midrule
Input parameters only             & --         & --         \\
Linear spectra only       & \checkmark & --         \\
Linear drive scalars only         & --         & \checkmark \\
Both linear inputs                 & \checkmark & \checkmark \\
\bottomrule
\end{tabular}
\label{tab:nl-variants}
\end{table}

To evaluate extrapolation toward more strongly driven conditions, the retained database is ranked according to the linear growth rate integrated over $k_y\leq1$. 
The upper 20\%, consisting of 132 cases, is assigned to a high-drive out-of-distribution (OOD) set, while the remaining 528 cases form the in-distribution set.
This partition deliberately places the cases with the strongest overall linear instability outside the training domain and provides a more demanding extrapolation test than a random split.

Performance within the in-distribution domain is evaluated using ten-fold out-of-fold predictions, repeated with four different data partitions while retaining the same in-distribution and high-drive cases. 
In each fold, 15\% of the remaining non-test cases are used as an internal validation set for checkpoint selection. 
The training initialization seed is fixed, and all four variants use the same fold and internal validation assignments within each repetition.
Within each repetition, each in-distribution case is predicted by the single fold model that used it as test data, excluding it from both training and validation.
For the high-drive OOD evaluation, predictions from all ten fold models within each repetition are combined into an ensemble.
For these OOD cases, the ensemble prediction of each positive output $y$ is the geometric mean
\begin{equation}
\widehat{y}_{\mathrm{ens}}
=\exp\left[\frac{1}{M}\sum_{m=1}^{M}
\log\bigl(\max(\widehat{y}_m,\epsilon)\bigr)\right],
\qquad M=10,
\label{eq:nl-ensemble-geometric}
\end{equation}
where $\widehat{y}_m$ is the prediction of fold model $m$ and $\epsilon=10^{-30}$ is a numerical floor used before taking the logarithm.
This calculation is applied to the reconstructed potential and heat flux spectra separately at each $k_y$, and also to the zonal flow amplitude and shearing rate.
The logarithmic average combines relative rather than absolute differences between predictions and gives less influence to large positive predictions than an arithmetic mean.
It is an aggregation rule and does not by itself provide a calibrated uncertainty interval.
Integrated spectral quantities and mode-summed totals are calculated from the ensemble spectra.

Performance metrics are calculated separately for each repetition and reported as their mean and sample standard deviation across the four data partitions.
Predictions are not pooled across repetitions, and the reported standard deviations describe partition sensitivity rather than the spread of the ten fold predictions.
The nonlinear comparison figures use the original data partition.

The spectral accuracy is evaluated separately over the low and high $k_y$ ranges. 
Linear-space $R^2$ is used for $k_y\leq1$, where the principal spectral amplitude is concentrated, while logarithmic $R^2$ is used for $k_y>1$ to resolve the weaker short-wavelength tail. 
The spectral $R^2$ measures prediction accuracy across simulation cases at each $k_y$, rather than agreement with the average spectral shape.
\revision{Logarithmic $R^2$ is used for the integrated spectra and zonal flow quantities.} 
Because $R^2$ is normalized by the target variance, its absolute values should not be compared directly between the in-distribution and high-drive sets.
We also report the normalized mean absolute error (NMAE), defined as $100\langle|\widehat{y}-y|\rangle/\langle|y|\rangle$ and evaluated in linear space.
For spectral bands, these averages include all cases and all $k_y$ points in the band.

\begin{table}[t]
\centering
\caption{Repeated ten-fold out-of-fold results for the 528 in-distribution cases. In each cell, the upper line gives $R^2$ and the lower line gives NMAE (\%), reported as the mean $\pm$ sample standard deviation across four data partitions. Metric definitions are given in the text. Boldface marks the largest mean $R^2$ in each row, including ties after rounding, and does not indicate statistical significance.}
\footnotesize
\setlength{\tabcolsep}{4pt}
\begin{tabular}{lcccc}
\toprule
& \shortstack{Input parameters\\only}
& \shortstack{Linear spectra\\only}
& \shortstack{Linear drive scalars\\only}
& \shortstack{Both linear\\inputs} \\
\midrule
$|\phi|$, low $k_y$ & \shortstack{$0.768\pm0.010$\\$(11.0\pm0.1)\,\%$} & \shortstack{$0.764\pm0.008$\\$(11.2\pm0.2)\,\%$} & \shortstack{$\mathbf{0.773\pm0.011}$\\$(10.8\pm0.2)\,\%$} & \shortstack{$0.769\pm0.007$\\$(11.0\pm0.1)\,\%$} \\[3pt]
$|\phi|$, high $k_y$ & \shortstack{$0.904\pm0.009$\\$(7.2\pm0.4)\,\%$} & \shortstack{$0.910\pm0.005$\\$(7.1\pm0.3)\,\%$} & \shortstack{$\mathbf{0.921\pm0.005}$\\$(6.6\pm0.3)\,\%$} & \shortstack{$0.916\pm0.005$\\$(7.0\pm0.2)\,\%$} \\[3pt]
$\int|\phi|\,d k_y$ & \shortstack{$0.911\pm0.008$\\$(5.6\pm0.3)\,\%$} & \shortstack{$0.907\pm0.006$\\$(5.6\pm0.2)\,\%$} & \shortstack{$\mathbf{0.919\pm0.008}$\\$(5.3\pm0.2)\,\%$} & \shortstack{$0.916\pm0.005$\\$(5.5\pm0.2)\,\%$} \\[3pt]
$Q$, low $k_y$ & \shortstack{$0.654\pm0.014$\\$(22.8\pm0.2)\,\%$} & \shortstack{$0.650\pm0.006$\\$(23.1\pm0.2)\,\%$} & \shortstack{$\mathbf{0.661\pm0.014}$\\$(22.5\pm0.2)\,\%$} & \shortstack{$0.658\pm0.009$\\$(22.7\pm0.3)\,\%$} \\[3pt]
$Q$, high $k_y$ & \shortstack{$0.931\pm0.008$\\$(12.2\pm0.8)\,\%$} & \shortstack{$0.932\pm0.005$\\$(11.7\pm0.5)\,\%$} & \shortstack{$\mathbf{0.941\pm0.003}$\\$(11.2\pm0.4)\,\%$} & \shortstack{$0.940\pm0.004$\\$(11.6\pm0.7)\,\%$} \\[3pt]
$\int Q\,d k_y$ & \shortstack{$0.908\pm0.005$\\$(11.8\pm0.5)\,\%$} & \shortstack{$0.902\pm0.005$\\$(11.7\pm0.3)\,\%$} & \shortstack{$\mathbf{0.914\pm0.009}$\\$(11.3\pm0.4)\,\%$} & \shortstack{$0.913\pm0.007$\\$(11.3\pm0.7)\,\%$} \\[3pt]
$|\phi_{\mathrm{ZF}}|$ & \shortstack{$0.893\pm0.005$\\$(10.9\pm0.4)\,\%$} & \shortstack{$0.894\pm0.007$\\$(10.8\pm0.3)\,\%$} & \shortstack{$\mathbf{0.903\pm0.007}$\\$(10.3\pm0.3)\,\%$} & \shortstack{$0.902\pm0.006$\\$(10.6\pm0.1)\,\%$} \\[3pt]
$|\omega_{E\times B}|$ & \shortstack{$0.895\pm0.031$\\$(8.6\pm0.9)\,\%$} & \shortstack{$0.915\pm0.003$\\$(7.8\pm0.2)\,\%$} & \shortstack{$0.923\pm0.004$\\$(7.5\pm0.3)\,\%$} & \shortstack{$\mathbf{0.927\pm0.008}$\\$(7.4\pm0.4)\,\%$} \\[3pt]
\bottomrule
\end{tabular}
\label{tab:nl-linear-ablation-id}
\end{table}

Table~\ref{tab:nl-linear-ablation-id} first compares the four variants using the in-distribution out-of-fold predictions. 
The four variants show comparable in-distribution performance, including the Input parameters only model.
The main limitation is the detailed spectral structure at low $k_y$, where the predictions are smoother than the nonlinear GENE spectra, which contain variations and secondary peaks.
For the Both linear inputs model, the low $k_y$ heat flux has an $R^2$ of 0.658 and an NMAE of 22.7\%, whereas its integrated magnitude has an $R^2$ of 0.913 and an NMAE of 11.3\%.
The model therefore reproduces the overall heat flux magnitude more accurately than its detailed distribution among the low $k_y$ modes.

\begin{table}[t]
\centering
\caption{High-drive results for the 132 out-of-distribution cases. Each repetition uses an ensemble of ten fold models. The upper and lower lines in each cell give $R^2$ and NMAE (\%), respectively, as the mean $\pm$ sample standard deviation across four data partitions. Metric definitions and boldface follow Table~\ref{tab:nl-linear-ablation-id}.}
\footnotesize
\setlength{\tabcolsep}{4pt}
\begin{tabular}{lcccc}
\toprule
& \shortstack{Input parameters\\only}
& \shortstack{Linear spectra\\only}
& \shortstack{Linear drive scalars\\only}
& \shortstack{Both linear\\inputs} \\
\midrule
$|\phi|$, low $k_y$ & \shortstack{$0.611\pm0.022$\\$(11.3\pm0.3)\,\%$} & \shortstack{$0.614\pm0.023$\\$(11.2\pm0.3)\,\%$} & \shortstack{$\mathbf{0.674\pm0.006}$\\$(10.2\pm0.2)\,\%$} & \shortstack{$0.672\pm0.008$\\$(10.2\pm0.2)\,\%$} \\[3pt]
$|\phi|$, high $k_y$ & \shortstack{$0.251\pm0.159$\\$(11.8\pm1.4)\,\%$} & \shortstack{$0.513\pm0.046$\\$(9.7\pm0.6)\,\%$} & \shortstack{$0.502\pm0.069$\\$(9.5\pm0.8)\,\%$} & \shortstack{$\mathbf{0.720\pm0.027}$\\$(7.3\pm0.4)\,\%$} \\[3pt]
$\int|\phi|\,d k_y$ & \shortstack{$0.529\pm0.068$\\$(8.3\pm0.6)\,\%$} & \shortstack{$0.573\pm0.043$\\$(7.8\pm0.5)\,\%$} & \shortstack{$0.765\pm0.022$\\$(5.8\pm0.3)\,\%$} & \shortstack{$\mathbf{0.771\pm0.014}$\\$(5.7\pm0.2)\,\%$} \\[3pt]
$Q$, low $k_y$ & \shortstack{$0.386\pm0.030$\\$(26.5\pm0.7)\,\%$} & \shortstack{$0.410\pm0.022$\\$(26.0\pm0.7)\,\%$} & \shortstack{$0.486\pm0.012$\\$(23.7\pm0.3)\,\%$} & \shortstack{$\mathbf{0.492\pm0.013}$\\$(23.8\pm0.5)\,\%$} \\[3pt]
$Q$, high $k_y$ & \shortstack{$0.288\pm0.128$\\$(26.6\pm2.3)\,\%$} & \shortstack{$0.458\pm0.090$\\$(23.4\pm2.2)\,\%$} & \shortstack{$0.607\pm0.012$\\$(20.2\pm0.4)\,\%$} & \shortstack{$\mathbf{0.723\pm0.029}$\\$(16.6\pm1.0)\,\%$} \\[3pt]
$\int Q\,d k_y$ & \shortstack{$0.448\pm0.069$\\$(20.2\pm1.2)\,\%$} & \shortstack{$0.503\pm0.060$\\$(19.1\pm1.3)\,\%$} & \shortstack{$\mathbf{0.712\pm0.023}$\\$(14.6\pm0.6)\,\%$} & \shortstack{$0.706\pm0.032$\\$(14.6\pm0.9)\,\%$} \\[3pt]
$|\phi_{\mathrm{ZF}}|$ & \shortstack{$0.725\pm0.036$\\$(16.5\pm1.1)\,\%$} & \shortstack{$0.791\pm0.026$\\$(14.2\pm1.0)\,\%$} & \shortstack{$0.901\pm0.020$\\$(9.5\pm1.1)\,\%$} & \shortstack{$\mathbf{0.904\pm0.013}$\\$(9.4\pm0.7)\,\%$} \\[3pt]
$|\omega_{E\times B}|$ & \shortstack{$0.484\pm0.093$\\$(11.5\pm1.1)\,\%$} & \shortstack{$0.672\pm0.035$\\$(8.8\pm0.5)\,\%$} & \shortstack{$0.854\pm0.013$\\$(5.8\pm0.3)\,\%$} & \shortstack{$\mathbf{0.871\pm0.009}$\\$(5.4\pm0.3)\,\%$} \\[3pt]
\bottomrule
\end{tabular}
\label{tab:nl-linear-ablation-ood}
\end{table}

The contribution of the linear inputs becomes more pronounced on the high-drive set, as shown in Table~\ref{tab:nl-linear-ablation-ood}.
Compared with the Input parameters only model, the Linear spectra only model improves the high $k_y$ potential and heat flux predictions, with smaller improvements in the low $k_y$ spectra and integrated amplitudes.
The Linear drive scalars only model gives better low $k_y$ and integrated-amplitude predictions than the Linear spectra only model, together with substantially better zonal flow amplitude and shearing rate predictions.
Compared with the Linear drive scalars only model, the Both linear inputs model further improves the high $k_y$ potential and heat flux predictions while retaining similar integrated-amplitude accuracy.
The zonal flow amplitude $R^2$ values are also similar (0.901 and 0.904), while the zonal flow shearing rate $R^2$ increases from 0.854 to 0.871.
The improvements in the high $k_y$ spectra and zonal flow shearing rate are observed in all four data partitions.
In the present architecture, the integrated amplitude heads receive the linear drive scalars directly, whereas the linear spectra affect these heads through the predicted zonal flow.
These results show complementary benefits from the two linear input paths in the present model on the high-drive set.

\begin{figure}[t]
\centering
\begin{overpic}[width=0.98\textwidth]{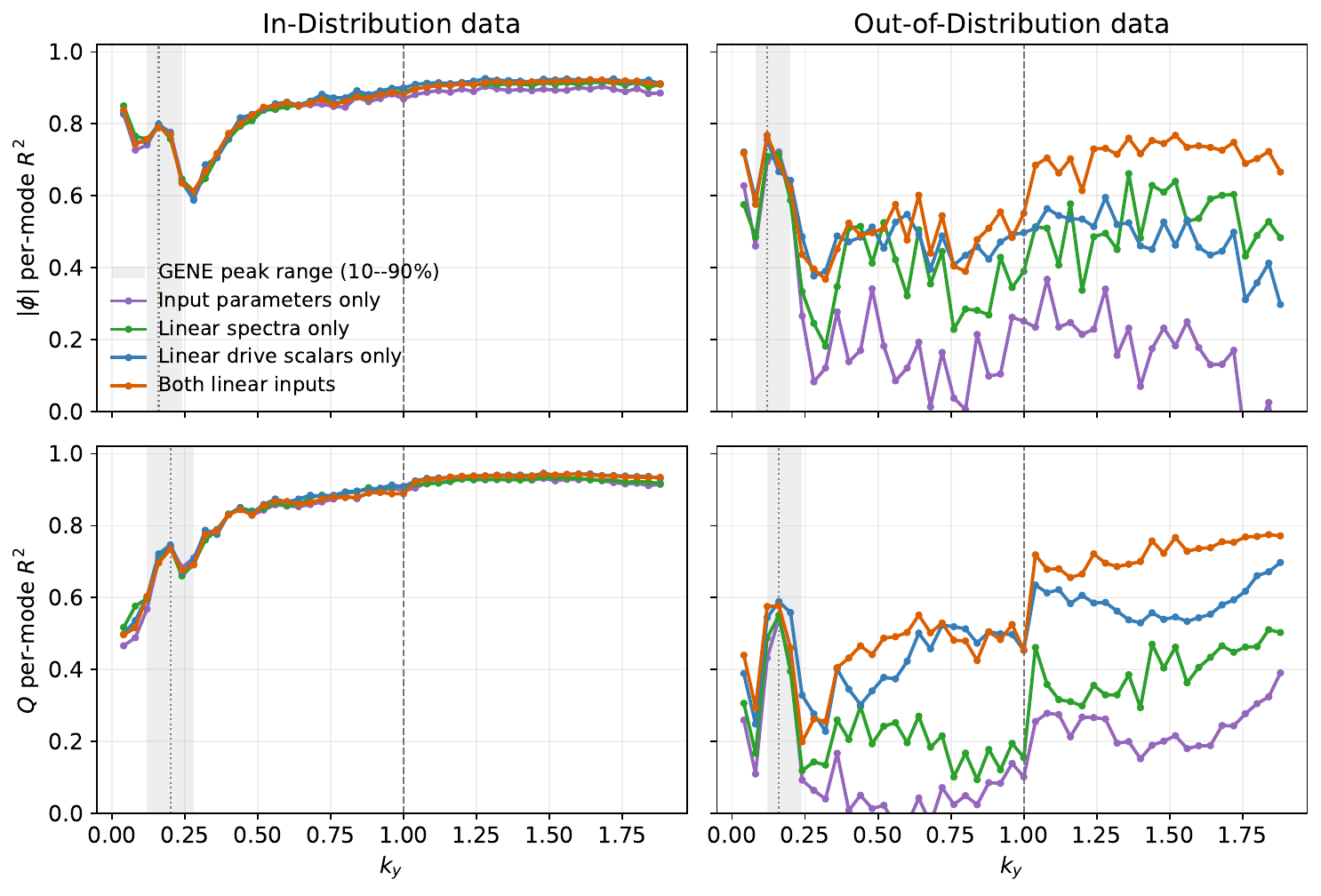}
\put(9,61){\textbf{a)}}
\put(56,61){\textbf{b)}}
\put(9,31){\textbf{c)}}
\put(56,31){\textbf{d)}}
\end{overpic}
\caption{Mode-resolved $R^2$ for potential (top) and heat flux (bottom) in the in-distribution (left) and high-drive OOD (right) sets. The dashed line at $k_y=1$ separates linear-space and logarithmic $R^2$ evaluations. The gray band shows the 10th--90th percentile range of GENE peak locations, and the dotted line marks the median.}
\label{fig:nl-ky-r2}
\end{figure}

Fig.~\ref{fig:nl-ky-r2} shows the $R^2$ at each $k_y$ for the in-distribution and OOD predictions to examine the spectral accuracy in more detail.
The gray band indicates the 10th--90th percentile range of the peak location in the GENE spectra, and the dotted line marks its median.
The vertical dashed line at $k_y=1$ separates the linear-space evaluation at $k_y\leq1$ from the logarithmic evaluation at $k_y>1$.
In distribution, the four models show similar performance.
Both the potential and heat flux predictions have relatively high $R^2$ near the principal spectral peak, but their accuracy decreases in nearby low $k_y$ regions.
For the potential, a local reduction appears above the peak-location range, around $k_y=0.24$--$0.28$, while the heat flux predictions also show reduced accuracy at the lowest sampled $k_y$ modes.
These results indicate that reproducing the spectral structure around the peak remains difficult even when the predictions near the peak are relatively accurate.
For the heat flux, the reduced accuracy at the lowest $k_y$ may be associated with difficulty reproducing the sharp decrease in the spectrum below the peak.

The differences between the models become clearer on the high-drive OOD set.
All four models show relatively high $R^2$ near the GENE peak-location range, followed by a pronounced reduction in the neighboring low $k_y$ region.
The reduction near $k_y\approx0.3$ is consistent with the same difficulty in reproducing the spectral structure around the peak as in the in-distribution cases.
The Linear drive scalars only and Both linear inputs models generally perform better in this region than the other two models.
For the potential, the Linear spectra only and Linear drive scalars only models show broadly comparable $R^2$ for $k_y>0.5$.
For the heat flux, however, the Linear drive scalars only model gives clearly higher $R^2$ than the Linear spectra only model over most of this range.
At high $k_y$, the Linear spectra only and Linear drive scalars only models both improve on the Input parameters only model, while the Both linear inputs model gives the highest $R^2$ for both potential and heat flux spectra.
This comparison shows that the benefit of combining the two linear input paths is most apparent in the high $k_y$ predictions on the high-drive set.

\begin{figure}[t]
\centering
\begin{overpic}[width=0.98\textwidth]{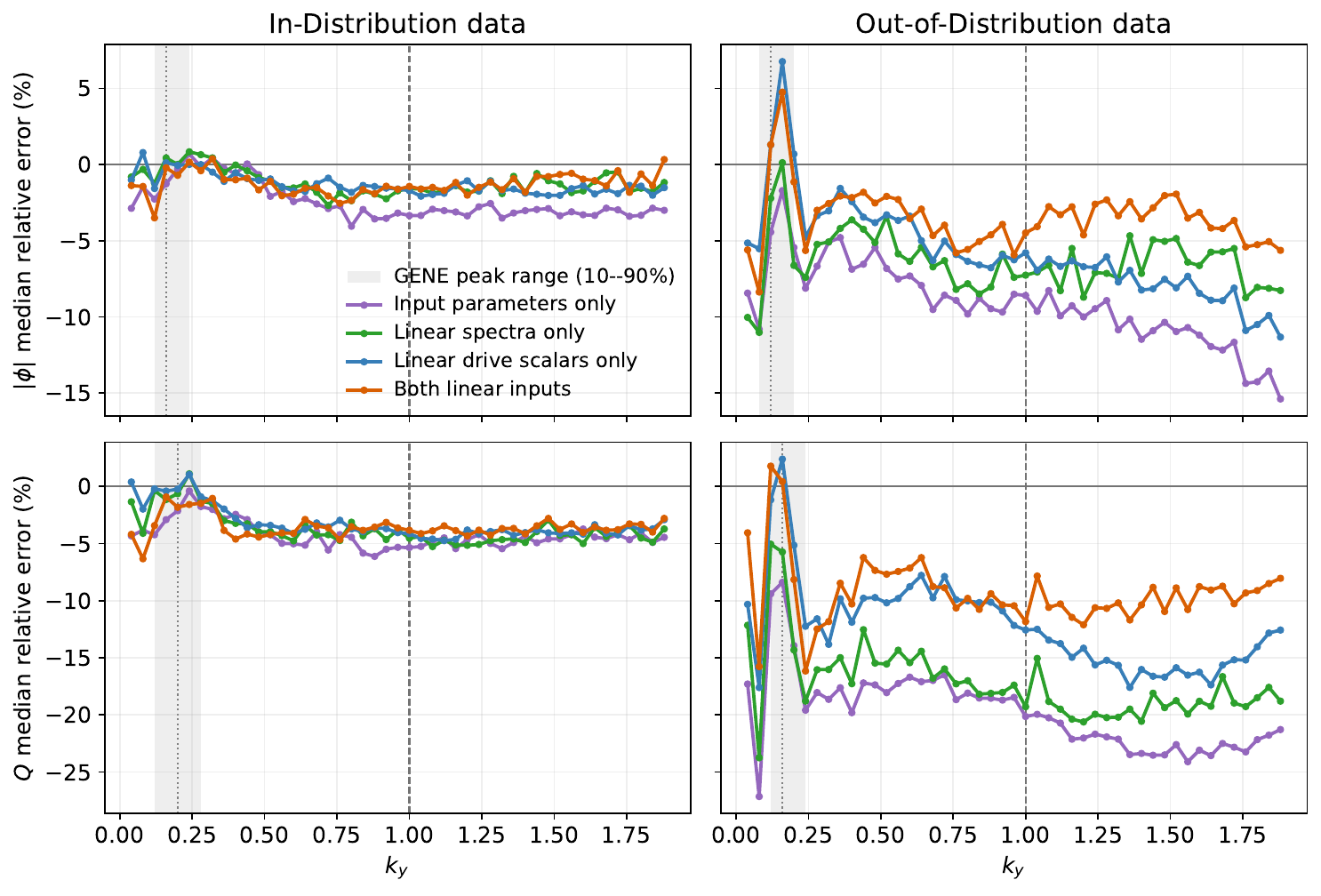}
\put(42,61){\textbf{a)}}
\put(92,61){\textbf{b)}}
\put(42,11){\textbf{c)}}
\put(92,27){\textbf{d)}}
\end{overpic}
\caption{Median signed relative errors for potential (top) and heat flux (bottom) in the in-distribution (left) and high-drive OOD (right) sets. Positive and negative values indicate overprediction and underprediction. The gray band and dotted line mark the 10th--90th percentile range and median of GENE peak locations.}
\label{fig:nl-ky-relative-error}
\end{figure}

Fig.~\ref{fig:nl-ky-relative-error} shows the median signed relative error at each $k_y$ to distinguish overprediction from underprediction.
A median error near zero does not imply small errors for individual cases.
The predictions tend to underestimate the spectra over much of the $k_y$ range, with stronger underprediction on the high-drive OOD set.
The Both linear inputs model reduces this tendency over much of the spectrum, although some underprediction remains.

The high $k_y$ predictions depend on both the spectral shape and the shared integrated amplitude.
The linear drive scalars can therefore improve the high $k_y$ predictions through a more accurate overall amplitude, even without linear spectra.
To assess the additional contribution of the linear spectra, we compare the Input parameters only model with the Linear spectra only model, and the Linear drive scalars only model with the Both linear inputs model.
Within each pair, the differences are relatively small at low $k_y$ but become more pronounced at high $k_y$, particularly on the high-drive set.
These comparisons are consistent with a contribution from the linear spectra to the short-wavelength predictions, although the errors also depend on the predicted overall amplitude.

\begin{figure}[t]
\centering
\begin{minipage}{0.98\textwidth}
\raggedright (a) Input parameters only\\
\includegraphics[width=\linewidth]{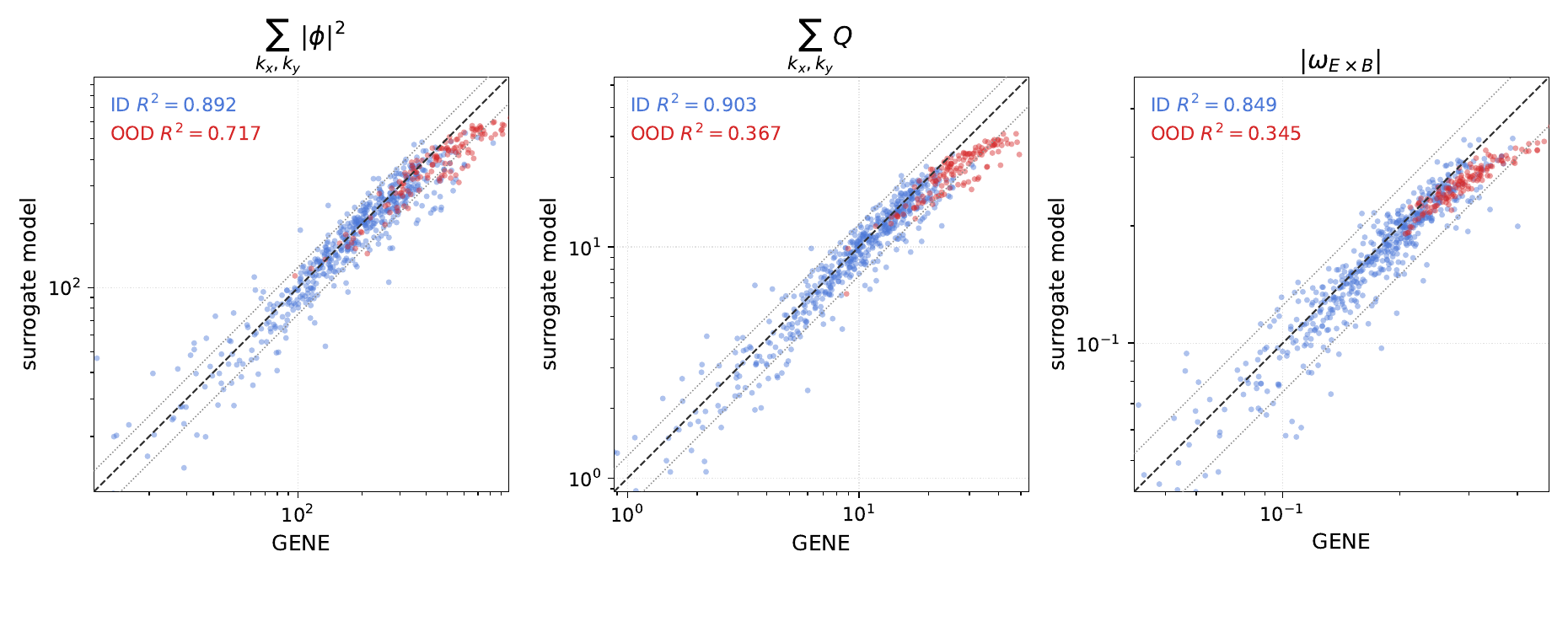}
\end{minipage}
\par\smallskip 
\begin{minipage}{0.98\textwidth}
\raggedright (b) Linear drive scalars only\\
\includegraphics[width=\linewidth]{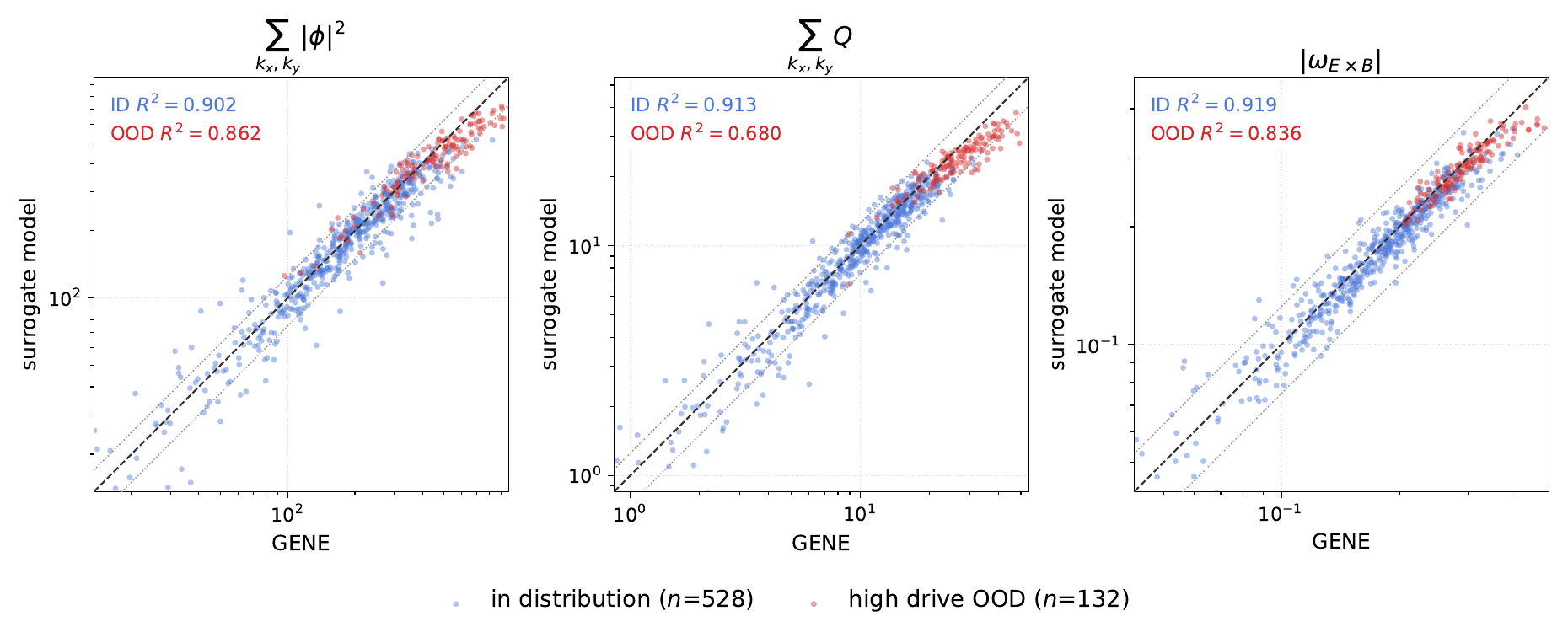}
\end{minipage}
\caption{Predicted total electrostatic energy, total heat flux, and zonal flow shearing rate for a) Input parameters only and b) Linear drive scalars only. The legend in b) applies to both panels. Dashed lines indicate equality, and dotted lines indicate deviations of $\pm25\%$. The displayed $R^2$ values are calculated in logarithmic space.}
\label{fig:nl-drive-scatter}
\end{figure}

To examine the effect of the linear drive scalars on the overall spectral magnitude and zonal flow predictions in more detail, Fig.~\ref{fig:nl-drive-scatter} compares the total electrostatic energy, total heat flux, and zonal flow shearing rate predicted by the Input parameters only and Linear drive scalars only models with the GENE results.
Both models show good agreement with the in-distribution results.
On the high-drive set, however, the Input parameters only model increasingly underestimates these quantities as the GENE values become larger.
The Linear drive scalars only model reduces this underprediction, although the heat flux remains less accurately predicted than the electrostatic energy and zonal flow shearing rate.
This remaining difficulty may partly reflect the dependence of heat flux on temperature fluctuation amplitudes and their cross-phases with the electrostatic potential, rather than on the potential amplitude alone.
These quantities are not explicitly supplied to the model and may not be sufficiently represented by the linear drive scalars.

\begin{figure}[t]
\centering
\includegraphics[width=0.98\textwidth]{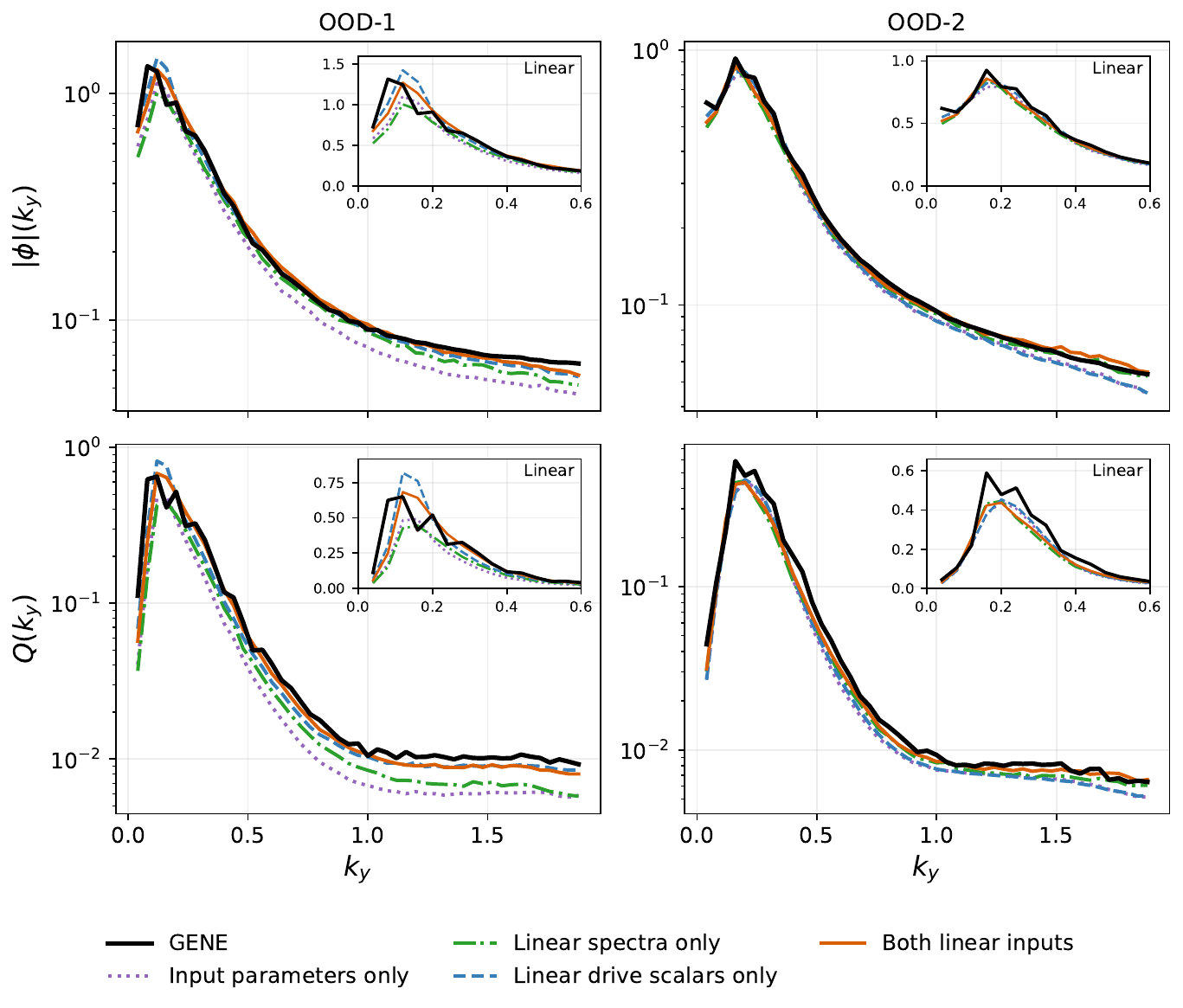}
\caption{Potential (top) and heat flux (bottom) spectra for two high-drive OOD cases. OOD-1 illustrates an advantage of the Linear drive scalars only model over the Linear spectra only model, whereas OOD-2 illustrates the opposite case. Black curves show GENE results, and colored curves show model predictions. Insets show the low $k_y$ peaks on linear axes. The input parameters $(R/L_{T_i},R/L_{T_e},R/L_n,q,\hat{s},\varepsilon)$ are approximately $(3.065,7.265,2.898,2.168,0.627,0.256)$ for OOD-1 and $(6.017,3.955,2.924,1.505,0.421,0.250)$ for OOD-2.}
\label{fig:nl-ood-spectra}
\end{figure}

To illustrate these differences in individual spectra, Fig.~\ref{fig:nl-ood-spectra} compares the predicted potential and heat flux spectra with the GENE results for two selected high-drive cases.
The logarithmic axes show the spectral tails, while the linear-scale insets resolve the low $k_y$ peaks.
OOD-1 is selected as a case where the Linear spectra only model has a larger heat flux spectral error than the other two models receiving linear inputs.
OOD-2 is selected as a case where the Linear drive scalars only model has the larger error.
The input parameters of both cases are listed in the figure caption.

In OOD-1, the Linear spectra only model predicts a lower spectral amplitude near the peak.
Its underprediction extends into the high $k_y$ range, suggesting that part of the tail error is associated with the overall spectral magnitude.
The models receiving the linear drive scalars better reproduce this magnitude.

In OOD-2, the linear-scale insets show similar peak amplitudes and low $k_y$ spectral predictions among the models receiving linear inputs.
Their predictions nevertheless separate at high $k_y$, where the Linear spectra only model follows the GENE spectra more closely than the Linear drive scalars only model.
This case illustrates the additional contribution of the linear spectra to the short-wavelength spectral structure, even when the low $k_y$ predictions are similar.

Taken together, these results show that lower-fidelity linear stability predictions can improve the predictions of a nonlinear turbulence surrogate model trained on a limited database.
The linear spectra and linear drive scalars provide complementary benefits in the present architecture, supporting the selection of the Both linear inputs model as the final nonlinear surrogate model.
These findings highlight the value of using physical understanding to guide both the choice of model inputs and the paths through which they contribute to the predicted quantities.

\clearpage

\section{Conclusion}
\label{sec:conclusion}

We have developed a two-stage surrogate model framework for ITG-TEM gyrokinetic turbulence using linear and nonlinear GENE simulations.
The linear surrogate model provides an efficient way to select unstable conditions for nonlinear simulations and supplies physical features for predicting the nonlinear response.
Its stability and mode classifiers are combined with global and mode-specific regressors, achieving held-out $R^2$ values of 0.9975 for the growth rate and 0.9730 for the real frequency, with 97.64\% classification accuracy.
For unstable cases with nonzero target magnitude, the median absolute relative errors are approximately 2\%.
The FNO-based nonlinear model then uses these linear predictions to learn saturated spectra and zonal flow quantities without imposing a quasilinear flux closure.

Four variants with different combinations of the linear spectra and linear drive scalars show comparable performance within the training domain.
Their differences become clearer for the high-drive cases, which are separated by the integrated linear growth rate.
Supplying the linear drive scalars directly to the scalar prediction heads reduces the underprediction of total nonzonal electrostatic energy, total heat flux, and zonal flow shearing rate.
In the architecture used for the main comparisons, adding the linear spectra through the FNO further improves the short-wavelength spectral predictions.
The main conclusion is that less expensive linear stability information can improve predictions learned from a limited nonlinear database, while its benefit depends on how it is used in the model.

The remaining limitations concern the detailed low $k_y$ spectral structure and underprediction on the high-drive set, particularly for the heat flux.
The principal spectral peak is generally located close to the GENE peak, but its height and localized multi-peak structures are not fully reproduced.
Integrated quantities are predicted more accurately than the detailed low $k_y$ spectral structure.
Performance near marginal stability is not established because such conditions are poorly represented in the nonlinear database.
Independent validation is also needed, since the high-drive set was examined during model development.

Future work will extend the nonlinear database and examine additional physical inputs, such as quasilinear flux predictions from QuaLiKiz~\cite{bourdelle2016qualikiz}.
After warm-up, the complete two-stage surrogate model takes approximately 20 ms per parameter point on an Apple M5 Pro CPU using six threads, including the ten-model nonlinear ensemble but excluding model loading.
This cost suggests potential applications in integrated plasma modeling and, after further validation, virtual fusion reactor calculations.

\ack{This research was supported by the National Research Foundation (NRF) funded by the Korean government (MSIT) (No. RS-2026-25545634).
This work was also supported by the R\&D Program through Korea Institute of Fusion Energy (KFE) funded by the Ministry of Science and ICT of the Republic of Korea (No. KFE-EN2641-12).
This work was also supported by the National Research Foundation, Singapore.
Computing resources were provided on the KFE computer, KAIROS, funded by the Ministry of Science and ICT of the Republic of Korea (No. KFE-EN2641-12).
Additional computing resources were provided by the National Supercomputing Centre, Singapore (https://www.nscc.sg).}

\appendix
\section{Output parameterizations and loss terms}
\label{sec:nl-training-details}

\revision{This appendix describes the output parameterizations and loss terms for the potential and heat flux spectra and zonal flow quantities.}
The target definitions and information paths are described in Sec.~\ref{sec:nl-model}, and the data partitions and evaluation protocol are described in Sec.~\ref{sec:nl-eval}.

The shape and amplitude heads produce logarithmic outputs. 
The output transformation of the shape head enforces a unit trapezoidal $k_y$ integral, and exponentiation ensures that both predictions are positive. 

The two heads are supervised separately. 
The shape head is trained against the normalized target spectrum using a mean squared error in logarithmic space, while the amplitude head is trained against the trapezoidal $k_y$ integral of the target using a relative mean squared error in linear space. 
This formulation allows the spectral distribution and its integrated magnitude to be learned as distinct target quantities.

The zonal flow amplitude is supervised using a mean squared error in logarithmic space, while the radial structure ratio is supervised using a relative mean squared error in linear space. 
Supervising the ratio directly prevents its loss from depending on the predicted zonal flow amplitude.

\revision{The loss contributions from these outputs and the spectral smoothness penalty are}
{\color{black}
\begin{align}
\mathcal{L}_{\mathrm{spec+ZF}}
={}&
\lambda_{S,\phi}\mathcal{L}_{S,\phi}
+\lambda_{T,\phi}\mathcal{L}_{T,\phi}
+\lambda_{S,Q}\mathcal{L}_{S,Q}
+\lambda_{T,Q}\mathcal{L}_{T,Q}
\nonumber\\
&+\lambda_{\mathrm{ZF},\phi}\mathcal{L}_{\mathrm{ZF},\phi}
+\lambda_{\mathrm{ZF},r}\mathcal{L}_{\mathrm{ZF},r}
+\lambda_{\mathrm{sm}}\mathcal{L}_{\mathrm{sm}}.
\label{eq:nl-spectral-zf-loss}
\end{align}
}
The smoothness term penalizes abrupt variations between adjacent $k_y$ locations in the logarithms of the predicted potential and heat flux spectra.
Its small weight is intended to suppress isolated numerical spikes without substantially modifying the resolved spectral structure. 

\revision{The weights of the terms in Eq.~\eqref{eq:nl-spectral-zf-loss} are listed in Table~\ref{tab:nl-loss-weights}.}
These weights were selected using a subset of the high-drive cases and then fixed across all variants.
Network parameters were trained exclusively on the in-distribution data.
The high-drive set was therefore used during model development and is not an independent final test set.

\begin{table}[htbp]
\centering
\caption{\revision{Weights of the spectral, zonal flow, and smoothness loss terms shown in Eq.~\eqref{eq:nl-spectral-zf-loss}, shared by all model variants.}}
\begin{tabular}{lccccccc}
\toprule
& $\lambda_{S,\phi}$ & $\lambda_{T,\phi}$
& $\lambda_{S,Q}$ & $\lambda_{T,Q}$
& $\lambda_{\mathrm{ZF},\phi}$ & $\lambda_{\mathrm{ZF},r}$
& $\lambda_{\mathrm{sm}}$ \\
\midrule
Weight & 0.296 & 1.158 & 0.797 & 0.553
& 0.235 & 0.572 & 0.001 \\
\bottomrule
\end{tabular}
\label{tab:nl-loss-weights}
\end{table}

\printbibliography

\end{document}